\documentclass[journal]{IEEEtran}
\usepackage{graphicx}
\usepackage{booktabs}
\usepackage{xcolor}
\usepackage{caption}
\usepackage{subcaption}
\usepackage{showlabels}
\usepackage{enumitem}
\usepackage{hyperref}
\usepackage{amsmath}
\usepackage[ruled,vlined]{algorithm2e}

\usepackage[noadjust]{cite}

\usepackage[utf8]{inputenc} 
\usepackage[T1]{fontenc}    
\usepackage{hyperref}       
\usepackage{url}            
\usepackage{booktabs}       
\usepackage{amsfonts}       
\usepackage{nicefrac}       
\usepackage{microtype}      
\usepackage{hyperref}
\usepackage{multirow, makecell}

\SetKwFunction{Range}{range}
\SetKwInput{KwInput}{Input}
\SetKwInput{KwParams}{Parameters}

\begin{document}
\title{BigSSL: Exploring the Frontier of\\
Large-Scale Semi-Supervised Learning\\
for Automatic Speech Recognition}

\author{Yu Zhang$^*$, Daniel S. Park$^*$, Wei Han$^*$, \\
James Qin, Anmol Gulati, Joel Shor, Aren Jansen, \\
Yuanzhong Xu, Yanping Huang, Shibo Wang, Zongwei Zhou, \\
Bo Li, Min Ma, William Chan, Jiahui Yu, Yongqiang Wang, \\
Liangliang Cao, Khe Chai Sim, Bhuvana Ramabhadran, \\
Tara N. Sainath, Fran\c{c}oise Beaufays, Zhifeng Chen, Quoc V. Le, Chung-Cheng Chiu, \\
Ruoming Pang$^\dagger$ and Yonghui Wu
\thanks{All authors, unless indicated otherwise, are affiliated with Google Inc.}
\thanks{$\dagger$Author is affiliated with Apple Inc. Work done while at Google.}
\thanks{*Equal contribution. Contact authors at \texttt{\scriptsize\{ngyuzh, danielspark, weihan\}@google.com}.}}

\maketitle

\begin{abstract}
We summarize the results of a host of efforts using giant automatic speech recognition (ASR) models pre-trained using large, diverse unlabeled datasets containing approximately a million hours of audio. We find that the combination of pre-training, self-training and scaling up model size greatly increases data efficiency, even for extremely large tasks with tens of thousands of hours of labeled data. In particular, on an ASR task with 34k hours of labeled data, by fine-tuning an 8 billion parameter pre-trained Conformer model we can match state-of-the-art (SoTA) performance with only 3\% of the training data and significantly improve SoTA with the full training set. We also report on the universal benefits gained from using big pre-trained and self-trained models for a large set of downstream tasks that cover a wide range of speech domains and span multiple orders of magnitudes of dataset sizes, including obtaining SoTA performance on many public benchmarks. In addition, we utilize the learned representation of pre-trained networks to achieve SoTA results on non-ASR tasks.
\end{abstract}

\section{Introduction}

Semi-supervised learning (SSL), which uses unlabeled data to enhance the performance of labeled tasks, has recently played a crucial part in improving public automatic speech recognition (ASR) benchmarks. A combination of pre-training \cite{hsu2018extracting, chung2018speech2vec, oord2018representation, chung2019autoregressive, chorowski2019unsupervised, schneider2019wav2vec, baevski2019vqwav2vec, ling2019deep, baevski2019effectiveness, riviere2020unsupervised, kawakami2020learning, wav2vec2, chen2021injecting} and self-training \cite{Zavaliagkos98utilizinguntranscribed,Lamel00lightlysupervised,Novotney2009,Thomas2013,li2019,kahn2019selftraining,synnaeve2019endtoend,parthasarathi2019,hsu2020selfsupervised,nstasr,xu2020iterative,chen2021semi} methods have been utilized to enable deep networks to push the state-of-the-art (SoTA) performance on public ASR datasets \cite{wav2vec2, nstasr, ssllimit}.

The dominant setting for semi-supervised learning has been in the domain of audio books. The Libri-Light dataset \cite{librilight}, which contains 60k hours of audio is by far the largest unlabeled semi-supervised dataset that has been used to improve the performance on LibriSpeech \cite{librispeech} and sub-sampled subsets thereof \cite{librilight, wav2vec2, nstasr, ssllimit, xu2021self}. Despite the success and exciting developments in this domain, this setting for semi-supervised learning is limited in a few aspects. First, the unsupervised data is tailored to the supervised task and pre-trained models on Libri-Light has shown limited generalization capacity to different domains in some instances \cite{speechstew}. Second, the Libri-Light dataset is not much bigger than industrial-scaled labeled datasets. Third, the supervised tasks considered are much smaller compared to practical tasks on which the performance of the network needs to be improved.

In this report, we study the utility of large models, with the parameter count ranging from 600M to 8B, pre-trained and self-trained on extremely large and diverse unlabeled datasets containing hundreds of thousands to a million hours of audio. More precisely, we construct:
\begin{itemize}
\item P-models: Models pre-trained on large unlabeled datasets.
\item PS-models: Models pre-trained and self-trained with large unlabeled datasets.
\end{itemize}

These models are in turn utilized to improve various labeled downstream tasks. We compile an extensive list of downstream tasks with audio data ranging from tens of hours to tens of thousands of hours across a wide variety of domains and languages. We focus on three different classes of downstream training methods:
\begin{itemize}
\item Training P-models on labeled datasets.
\item Downstream self-training with P-models.
\item Fine-tuning with PS-models.
\end{itemize}
The first two training methods only employ unlabeled data in addition to the labeled data of the downstream task. Meanwhile, PS-models are self-trained upstream on a large amount of additional pseudo-labeled data.

For the rest of the section, we highlight key findings, present an overview of the paper and comment on related work.

\subsection{Key Findings}

\begin{figure*}[t!]
\centering
\includegraphics[width=0.97\textwidth]{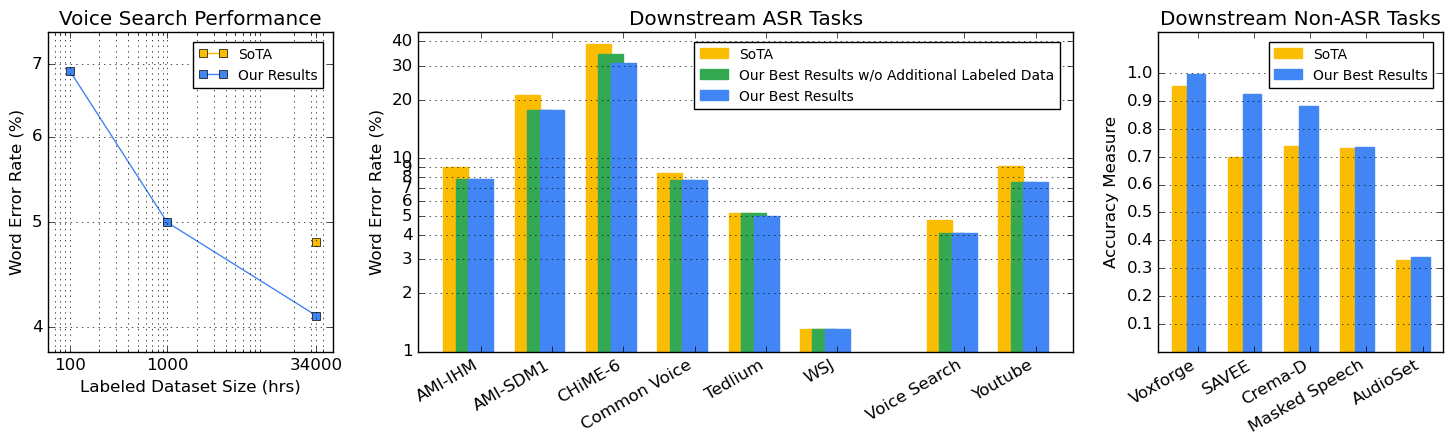}
\caption{\textbf{(Left)}$^\dagger$ WERs (\%) of P-models trained on subsets of Voice Search. \textbf{(Middle)}$^\dagger$ SoTA results on public and non-public ASR benchmarks. \textbf{(Right)} SoTA results on public audio classification tasks. Accuracy measures presented are classification accuracy for Voxforge/SAVEE/Crema-D, unweighted average recall for Masked Speech, and mean average precision (mAP) for AudioSet. SoTA for AudioSet is selected from non-ensembled results using audio data exclusively. $^\dagger$Axes are log-scaled.}
\label{f:summary}
\end{figure*}

\textbf{SSL + Large Models = Labeled Data Efficiency: } By scaling up the model size and utilizing semi-supervised learning techniques with a large amount of unlabeled data, we vastly improve labeled data efficiency. The first panel of Figure \ref{f:summary} shows the performance we achieve, without the use of additional labeled data, by training our models on 100h, 1000h subsets of the 34kh training set of the English (US) Voice Search task (VS). We obtain comparable results with reported SoTA performance \cite{li2021better} by using only 3\% of the labeled data.

\textbf{SoTA results for downstream ASR tasks: } We exceed or match state-of-the-art results by fine-tuning the pre-trained models on a wide variety of downstream ASR tasks, as summarized in the second panel of Figure \ref{f:summary}. Results for all downstream ASR tasks studied are collected in Section \ref{s:downstreamasr}.

\textbf{SoTA results for downstream non-ASR tasks: } We have trained shallow classifiers on top of features derived from pre-trained large encoders for audio classification. By doing so, we are able to achieve SoTA on multiple public benchmarks as presented in the last panel of Figure \ref{f:summary}. Complete results are presented in Section \ref{s:downstreamnonasr}.

\begin{figure}[h!]
\centering
\includegraphics[width=0.85\columnwidth]{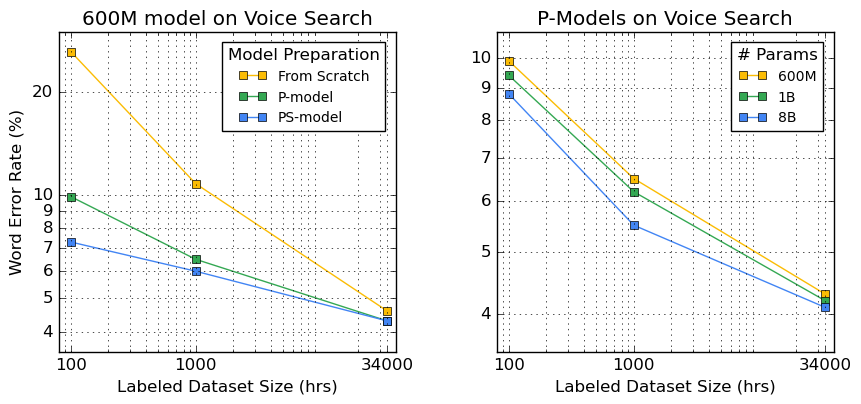}
\caption{WERs (\%) of models trained on subsets of Voice Search. On the left, we show the performance of the 600M-parameter model (the "Conformer XL") with varying preparation methods, while on the right, we report that of P-models of varying sizes.}
\label{f:returns}
\end{figure}

\textbf{Benefits of using SSL + Large Models are smaller for bigger downstream tasks, but are still significant: } The gains achieved by increasing model size, pre-training and self-training have diminishing returns with larger labeled dataset size as is shown in Figure \ref{f:returns}. Nevertheless, we are able to observe meaningful gains for industrial-scaled tasks.

\subsection{Outline}

The outline of this report is as follows:

\textbf{Methods:} We use the Conformer \cite{conformer} architecture as the speech encoders. We train 600M, 1B and 8B-parameter Conformer encoders using wav2vec 2.0 pre-training \cite{wav2vec2}, and self-train and/or fine-tune RNN-T \cite{rnnt} or CTC \cite{ctc} models having this encoder.

\textbf{Model Preparation with YouTube:} We use YouTube-based large-scale data to pre-train or self-train the Conformer models. A 1M-hour unlabeled dataset, which we denote YT-U, is used for pre-training, while a 500k-hour filtered unlabeled dataset, denoted YT-T, is used for self-training. P- and PS-models trained using these datasets are constructed.

\textbf{ASR Tasks:} We fine-tune the P- and PS-models on various ASR tasks and improve their performance. We are able to match existing benchmarks on the Voice Search task using 3\% of the full data, and significantly improve the performance of the full task by pre-training. We are also able to achieve SoTA/near-SoTA performance on YouTube and public datasets.

\textbf{Experiments with Voice Search:} We fine-tune the P- and PS-models on the 34k-hour English (US) Voice Search dataset \cite{li2021scaling} and its 100h, 1000h subsets. We run control experiments studying the effect of the labeled dataset size, model size, pre-training, upstream and downstream self-training. Cross-lingual benefits of pre-training are also explored.

\textbf{Non-ASR Tasks:} We use features derived from the intermediate layers of the P-models for non-ASR tasks. We achieve SoTA performance for multiple tasks within the non-semantic speech (NOSS) benchmark \cite{trill} by directly using these features with linear models only. For the AudioSet benchmark \cite{gemmeke2017audio}, which involves a wide variety of non-speech sounds, we find intermediate Conformer layers pre-trained on the non-labeled native dataset, rather than YT-U, to yield SoTA results.

\textbf{Discussions and Future Directions:} We comment on some noteworthy observations and discuss possible future directions.

Despite the unifying theme of employing SSL to train large models, the particulars of the experiments are varied, and not all experimental options are exhaustively explored in our control experiments. This is due to the fact that the tasks explored in this work have different pre-existing set-ups with limited budgets for experimentation. We describe the important elements of each experiment in the corresponding section and provide additional details in the appendix.

\subsection{Related Work}

Our work is an extension of a host of recent research efforts \cite{librilight, wav2vec2, nstasr, ssllimit, xu2021self} that have studied semi-supervised learning \cite{scudder1965probability, yarowsky1995unsupervised, riloff2003learning} for ASR in the context of deep-learning. Our main contribution is that we have scaled up pre-training and self-training both in terms of model size (8 billion parameters), unlabeled dataset size (a million hours of audio) and labeled dataset size (34 thousand hours of audio) and:
\begin{itemize}
\item Conducted a systematic study of the effect of pre-training, upstream/downstream self-training and model size on downstream tasks of varying sizes.
\item Fine-tuned the prepared models on seven public and six industrial ASR datasets spanning multiple speech domains, languages and accents.
\item Studied the utility of the pre-trained representations by using it for downstream non-ASR tasks.
\end{itemize}

We have used Conformers \cite{conformer} as the encoder architecture in this work. The pre-training method used in this work is based on wav2vec 2.0 \cite{wav2vec2}, while the self-training method is based on noisy student training \cite{nst} using SpecAugment \cite{specaugment, specaugment2}. These methods have been employed in the context of LibriSpeech in \cite{ssllimit}. There is extensive literature on pre-training \cite{hsu2018extracting, chung2018speech2vec, oord2018representation, chung2019autoregressive, chorowski2019unsupervised, schneider2019wav2vec, baevski2019vqwav2vec, ling2019deep, baevski2019effectiveness, riviere2020unsupervised, kawakami2020learning, wav2vec2, chen2021injecting} and self-training \cite{Zavaliagkos98utilizinguntranscribed,Lamel00lightlysupervised,Novotney2009,Thomas2013,li2019,kahn2019selftraining,synnaeve2019endtoend,parthasarathi2019,hsu2020selfsupervised,nstasr,xu2020iterative,chen2021semi} in ASR, a subset of which we list in the bibliography. Methods for improving the performance of streaming ASR models using models with future context have been studied in \cite{serdyuk2017twin,kim2017improved,ravanelli2018twin,takashima2018investigation,kurata2018improved,kurata2020knowledge,yu2020universal, doutre-icassp21, doutre-interspeech21}. Multi-domain training, which has been used for training on multiple public datasets in this report, has been studied in the context of ASR in \cite{kaldi-multi-en, multidomain, likhomanenko2020rethinking, speechstew, li2021scaling}.

Giant models have been studied predominantly in the context of natural language processing \cite{bert, radford2019language}. Various methods have been employed for making giant models practically trainable \cite{shazeer2018mesh, shoeybi2019megatron, rajbhandari2020zero, xu2020automatic, gshard-arxive}. We have used the GShard~\cite{gshard-arxive} framework with the GSPMD backend~\cite{gspmd} to scale our ASR models up to 8B parameters.

\section{Methods}

The methods employed for experiments in this report largely follow that of \cite{ssllimit}. We review the key components here for completeness, while more details for each experiment can be found in the appendix.

\subsection{Model Architecture: Conformer}
\label{ss:conformer}

\begin{figure}[h!]
\centering
\includegraphics[width=0.95\columnwidth]{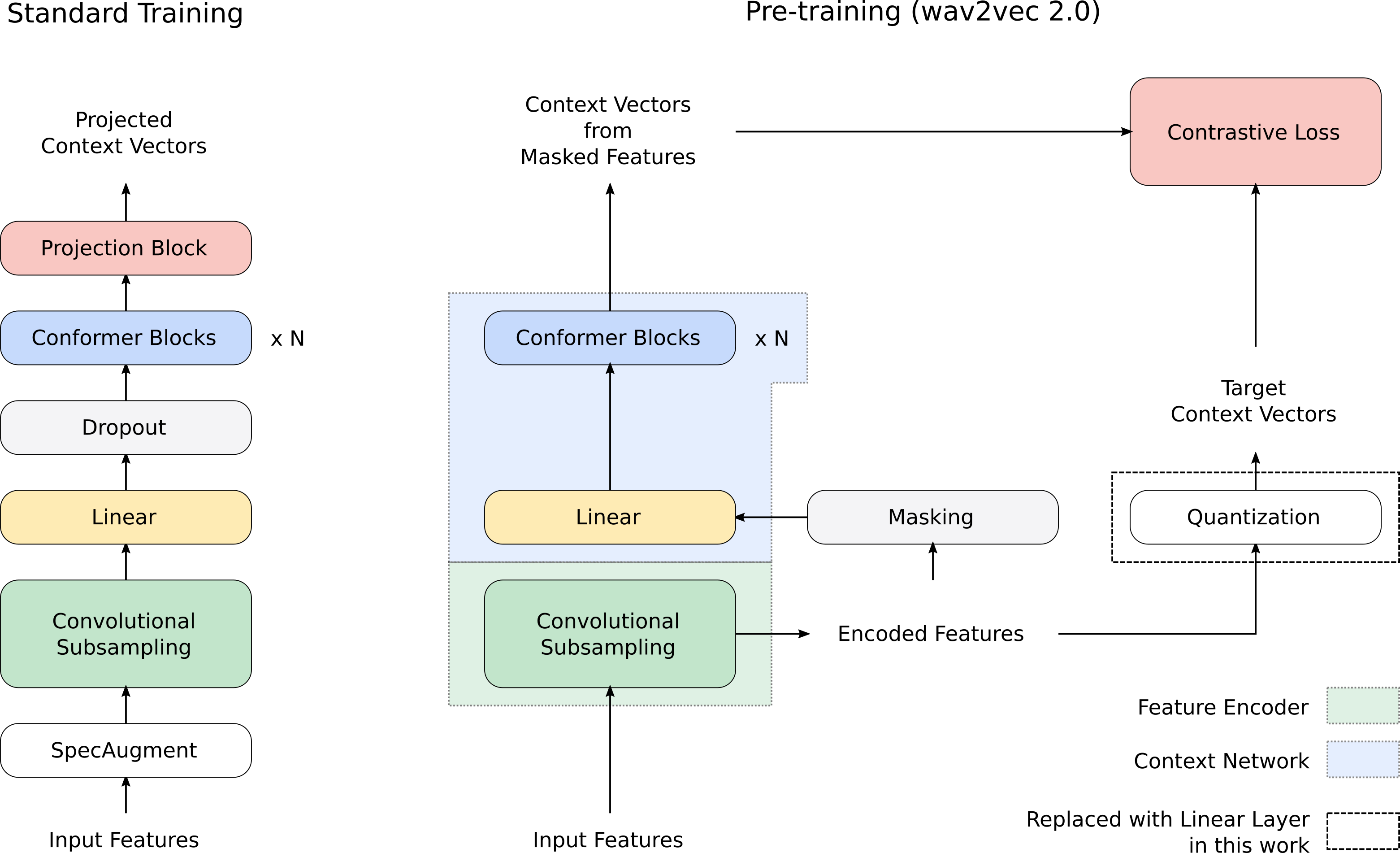}
\caption{The Conformer encoder and wav2vec 2.0 pre-training.}
\label{f:conformer}
\end{figure}

We use the Conformer \cite{conformer}, the convolution-augmented transformer, as the encoder network for our ASR models. The key component of the Conformer is the Conformer block, which consists of attention, feed-forward and convolutional modules \cite{conformer}. As depicted in the left panel of Figure \ref{f:conformer}, the input mel-log spectrogram to the network is subject to convolutional sub-sampling, after which a series of Conformer blocks and a projection layer are applied to obtain the final features.

These features are either used as input to an RNN transducer \cite{rnnt} along with a 2-layer LSTM decoder, or used as input for a connectionist temporal classification (CTC) model \cite{ctc} after an additional projection layer.

We consider three models with 600M, 1B and 8B parameters in this work, the particulars of which are listed in Table \ref{t:cparams}. The convolutional kernal size for all these models are set to 5. Following the notation of \cite{conformer,ssllimit}, we denote the three models, Conformer XL, Conformer XXL and Conformer G. We use relative attention \cite{dai2019transformer} for these models.

\begin{table}[h!]
  \caption{Conformer model parameters.}
  \label{t:cparams}
  \centering
  \resizebox{0.98\columnwidth}{!}{%
  \begin{tabular}{lccccc}
    \toprule
    Model & \# Params (B) & \# Layers & Dimension & Att. Heads\\
    \midrule
    Conformer XL & 0.6 & 24 & 1024 & 8 \\
    Conformer XXL & 1.0 & 42 & 1024 & 8 \\
    Conformer G & 8.0 & 36 & 3072 & 16 \\
    \bottomrule
  \end{tabular}}
\end{table}

\subsection{Pre-training: Wav2vec 2.0}

To pre-train the Conformer encoder network, we employ wav2vec 2.0 \cite{wav2vec2} training, as depicted in Figure \ref{f:conformer}. After first extracting encoded features from the convolutional sub-sampling layer of the network, we pass the features through the rest of the Conformer model after masking them to generate context vectors. These context vectors are trained to agree with the target context vectors, obtained by applying a linear layer to the initial encoded features, by a contrastive loss \cite{oord2018representation}. The convolutional subsampling layer consists of two 2D convolutional layers applied with strides (2, 2).

\subsection{Self-training: Noisy Student Training}

Noisy student training (NST) \cite{nst, nstasr} is a self-training method where a teacher model generates pseudo-labels for a large unlabeled dataset, which is in turn used to train a student model with augmentation.

The experimental set-up in this report differs from previous works including \cite{nstasr, ssllimit}, where the teacher model has been fused with a language model to generate better labels. In this work, our unlabeled dataset being very large, we choose to carry out inference without language model fusion, as the inference speed of the fused model is significantly slower than that of the ASR model on its own.

Unlike the fused model, the loss computed by the ASR model has a straightforward interpretation as a confidence measure---we thus can use a simple confidence-per-word measure to filter the teacher-generated transcripts. In our experiments, we either choose to retain the entire pseudo-labeled set, or filter 50\% of the utterances based on confidence-per-word.

For some experimental set-ups, we choose to use a small model that has already been trained on a labeled dataset for a given task as a teacher model, rather than using a scaled-up family of models as commonly done in the literature \cite{nst, nstasr}. We have also chosen to only proceed with one generation of NST training when it is employed to observe its effects, rather than optimize performance by going through many generations.

In conclusion, given a labeled dataset $S$, an unlabeled dataset $U$, our NST procedure is as follows:
\begin{enumerate}
\item Use teacher model $T$ to generated pseudo-labeled dataset $T(U)$. $T$ may be a model trained only with $S$, or one that has been pre-trained or self-trained.
\item (Optional) Filter $T(U)$ using confidence-per-word.
\item (Optional) Mix dataset $T(U)$ and $S$ into new training set.
\item Fine-tune new pre-trained model $M$ with augmentation on training set.
\end{enumerate}

\subsection{Gshard/GSPMD: Making 8B-parameter Models Trainable and Efficient}
We use the GShard~\cite{gshard-arxive} framework with the GSPMD backend~\cite{gspmd} to train the 8B model on Cloud TPUs. In particular, the GSPMD-style pipeline parallelism works very well. It is because 1) the model has many layers but each layer is not very large, which makes pipeline parallelism more efficient than other forms of model parallelism in terms of inter-device communication cost; 2) each pipeline stage is smaller than the full program, reducing the overall compilation/startup time; 3) we only need to pipeline the Conformer blocks, and GSPMD allows us to conveniently switch to data parallelism for the layers before and after.

\subsection{Training Details}
\label{ss:details}

\textbf{Data Processing: } The audio in this work has been uniformly sampled to 16 KHz quality---any audio with a different native sampling rate is either up-sampled or down-sampled. The audio is featurized into 80-dimensional log-mel filterbank coefficients. Two tokenization schemes are used for the transcripts: graphemes or word-piece models (WPMs) \cite{wpm}, the details of which differ for each experiment.

\textbf{Pre-training: } The masking parameters for wav2vec 2.0 pre-training are taken from \cite{wav2vec2}, where the starting point for the mask is chosen randomly with probability 0.065, and the mask size is set to 10 steps. The transformer learning rate schedule (section 5.3 of \cite{vaswani2017attention}), parameterized by the peak learning rate and the warm-up steps is used universally. The Conformer XL is trained using Adam optimization with exponential moving averaging (EMA) with decay rate 0.9999. The XXL/G models are trained with Adafactor \cite{adafactor} optimization. These model do not use exponential moving averages during pre-training.

\textbf{Training with Labels: } In this work, we encounter three kinds of initialization conditions with labeled training, one where the entire network is randomly initialized, one where only the encoder portion has been trained with wav2vec 2.0 pre-training and the rest of the network is randomly initialized, and one where the entire network has been trained in some fashion. To handle all three cases, we use two separate optimizers for the encoder parameters and the decoder parameters of the network during labeled training \cite{ssllimit}. As with pre-training the Conformer XL is trained with Adam optimization, while the Conformer XXL and G are trained with Adafactor optimization. All networks are trained using EMA with decay rate 0.9999 for labeled training. The learning rate schedule for both the decoder and encoder optimizers for the XL/XXL models are transformer schedules parameterized by peak learning rate and warm-up steps, while for the G model, we use a constant schedule with a linear warm-up phase. The batch size, learning rate and warm-up steps are adjusted for the downstream tasks. We use the standard adaptive SpecAugment \cite{specaugment, specaugment2} policy with two frequency masks with size parameter $F= 27$, and ten time masks with maximum time-mask ratio $p_S= 0.05$ to augment the input spectrogram.

\section{Model Preparation with YouTube}

We describe the procedure we use to prepare large Conformers using a large unlabeled dataset for downstream tasks in this section. The large unlabeled datasets that form the basis of our studies come from YouTube videos. The pre-trained and self-trained Conformers used to further train on downstream tasks for the rest of the paper are summarized in Section \ref{ss:ytmodels}.

\subsection{Data}
\label{ss:ytdata}

We collect three datasets based on YouTube, that is used to pre-train, self-train and train our models for downstream tasks, and tasks native to YouTube:
\begin{itemize}
\item YT-L: 350k hours of segmented, weakly-labeled audio, combined with 1000 hours of labeled audio.
\item YT-T: 500k hours of segmented, pseudo-labeled audio.
\item YT-U: 900k hours of segmented, unlabeled audio.
\end{itemize}
In constructing and pseudo-labeling the datasets, an important role is played by an RNN-T model with a bi-directional LSTM encoder. This 100M-parameter model will be referred to as the "YT (LSTM) teacher model" throughout this section.%
\footnote{See section 3.1 of \cite{chiu2021rnn} for details on the architecture/training of this model.} The audio segmentation is done by conducting inference with this teacher model, which is used to identify the speech boundaries. The model is further used to pseudo-label the YT-T dataset.

\textbf{YT-L: } YT-L is a combination of a weakly-labeled dataset whose method of construction has been elaborated on in \cite{ytwl} and an additional 1000 hours of labeled audio. The weakly labeled portion of the dataset is based on audio from videos that have user-uploaded transcripts, where "islands" of the audio and the transcripts are selected where the transcripts are thought to well-represent the audio. This is done by first force-aligning the transcripts and audio and finding islands of high confidence using a pre-existing acoustic model. 350k hours of audio with transcripts are obtained this way. 

\textbf{YT-T: } YT-T is a dataset also with audio from videos that have user-uploaded transcripts. These videos are further segmented using the YT teacher model, and the non-speech segments are removed, leaving 500k hours of audio. The user-provided transcripts of this dataset, however, are discarded and are {\it not} used for training. Instead, we choose to generate pseudo-labels on this dataset using the YT teacher model trained on YT-L when we use it for labeled training.

\textbf{YT-U: } YT-U is built by first randomly collecting 3 million hours of audio from "speech-heavy" YouTube videos, including lectures, news and interviews, filtered by language. The 3 million hours of audio is then further segmented by the YT teacher model. The non-speech segments identified by the YT teacher model are removed to yield approximately a million hours of unlabeled audio data.

The test set for the YT domain is generated by hand-transcribing popular videos from YouTube with 11 hrs of audio with lengths 2 - 10 min.

\begin{figure}[h!]
  \centering
  \includegraphics[width=0.75\columnwidth]{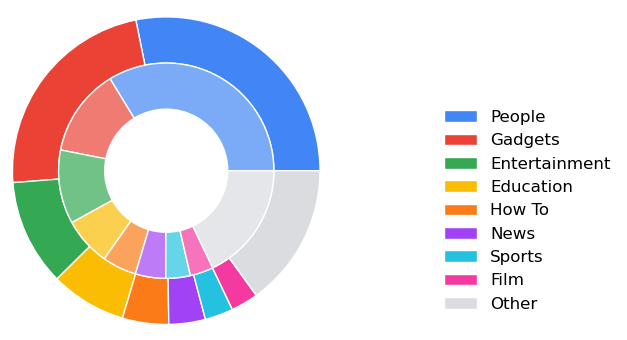}
  \caption{Video categories by length (outer) and number (inner).}
  \label{f:video}
\end{figure}

Besides the obvious advantage of having a very large amount of audio available, these YouTube-based datasets have an extremely wide range of sub-domains \cite{multidomain, ytwl}, plotted in Figure \ref{f:video} for YT-U. This variety will prove to be beneficial in various downstream tasks.

\subsection{Pre-trained and Self-trained Models}
\label{ss:ytmodels}

We produce Conformer models that have been pre-trained and self-trained with YouTube-based data that will be employed for downstream tasks in the rest of the paper. The notation
\begin{equation}
\textbf{Conformer\{Size\}-\{Decoder\}-\{Preparation\}} \nonumber
\end{equation}
is used to denote a Conformer of size "Size" with decoder-type "Decoder"  prepared with method "Preparation." The following options are available for each parameter:
\begin{itemize}
\item \textbf{Size}: XL, XXL or G.
\item \textbf{Decoder}: CTC or RNNT.
\item \textbf{Preparation}: Null (no preparation), P (pre-trained) or PS (pre-trained and self-trained).
\end{itemize}

Pre-training is done via wav2vec 2.0 with YT-U, the details of which are presented in Section \ref{ss:details}. This process exclusively prepares the encoder portion of the model.

Meanwhile, the PS-model is produced by taking the model pre-trained with YT-U and training it on the pseudo-labeled dataset YT-T. While the PS-model never sees the labeled dataset YT-L, it implicitly uses the information, since YT-T has been pseudo-labeled by the YT teacher model trained on YT-L. A 4k-WPM model is used to tokenize the text for PS-models. Self-training produces a strong upstream model that can be fine-tuned on smaller tasks. The performance of the PS-models on the YT test set is presented in Table \ref{t:ytmain}.

\section{ASR Tasks}
\label{s:downstreamasr}

In this section, we fine-tune the models pre-trained and self-trained with YouTube data on downstream ASR tasks. We compare the performance of our fine-tuned models with existing benchmarks, and show that they are able to improve state-of-the-art results on benchmarks spanning a wide range of dataset sizes and domains. We first present the effect of pre-training and self-training on the native YouTube task, and move on to presenting the improvements on the English (US) Voice Search task we were able to achieve both in terms of performance and efficiency. We then present our results on two public tasks, SpeechStew \cite{speechstew} and CHiME-6 \cite{chime6} and compare them to the current SoTA benchmarks. We also apply our methods to a non-public Telephony task and show that we are able to improve the performance of a streaming model by using a fine-tuned Conformer PS-model as a teacher.

\subsection{YouTube}

\textbf{Data:}
YT-L, which is partially labeled and partially weakly-labeled, is used as the supervised dataset for the YouTube task. YT-T is utilized as the unlabeled dataset for self-training.

\textbf{Results:}
We have presented our results of training pre-trained Conformer CTC and RNN-T models with labeled YT data against existing baselines in Table \ref{t:ytmain}. For self-training, the student Conformer model is trained with YT-T pseudo-labeled by the YT LSTM teacher model. The pseudo-labeled data is neither filtered nor mixed with YT-L for training the student. Quite remarkably, we find that training our models entirely with machine-generated transcripts turn out to show better performance than with the default labeled dataset.

\begin{table}[h!]
  \caption{WER (\%) on test sets after pre-training and self-training Conformer CTC and RNN-T models. We denote the models trained with pre-training + self-training as PS-models throughout the paper (see section \ref{ss:ytmodels}.)}
  \label{t:ytmain}
  \centering
  \resizebox{0.98\columnwidth}{!}{%
  \begin{tabular}{lrclr}
    \toprule
    Model & YT-test &&Model & YT-test \\
    \midrule
    SoTA \cite{chiu2021rnn} & 9.1 \\
    YT LSTM Teacher & 8.4 \\
    \midrule
    ConformerXL-CTC-P & 8.6 && ConformerXXL-CTC-P & 8.5\\
    + Self-training (PS) & 7.9 && + Self-training (PS) & \textbf{7.5} \\
    \midrule
    ConformerXL-RNNT-P & && ConformerXXL-RNNT-P &\\
    + Self-training (PS) & 7.8 && + Self-training (PS) & 7.8 \\
    \bottomrule
  \end{tabular}}
\end{table}

We note that the performance of the XXL models do not significantly improve beyond that of the XL models on YouTube. We hypothesize that this could due to the high level of noise of the labels of YT-L and YT-T.

\subsection{English (US) Voice Search}

\textbf{Data:}
The English (US) Voice Search (VS) dataset contains 34k hours of labeled voice search audio \cite{li2021scaling}. To test our ability to improve data efficiency, we construct random 1000 and 100 hour subsets, which we denote VS-1000h and VS-100h. For this data, we are able to utilize a 128M-parameter Conformer language model trained on a relevant text corpus, which is used for improving the ASR model performance via shallow-fusion \cite{shallowfusion} for RNN-T models. Further information on the dataset can be found in Section \ref{ss:searchdata}.

\textbf{Overview:}
We have conducted a systematic study of the effect of model size, labeled dataset size, pre-training, language model fusion, upstream and downstream self-training for training Conformers on VS, which we present in Section \ref{s:VS}.

\textbf{Main Results:}
Our best results for each subset of Voice Search are obtained by training the ConformerG-P. For the 100h subset, we have applied NST and LM fusion to obtain our best result, while for the 1000h subset, LM fusion is sufficient (see Section \ref{ss:searchnst}). For the full dataset, neither NST nor LM fusion pushed the performance further. The best WERs obtained have been presented in Table \ref{t:vsmain} and plotted in the first panel of Figure \ref{f:summary} in the introduction.

\begin{table}[h!]
  \caption{English (US) Voice Search test WERs (\%) from training on 100h, 1000h and 34kh subsets of VS.}
  \label{t:vsmain}
  \centering
  \resizebox{0.75\columnwidth}{!}{%
  \begin{tabular}{lccc}
    \toprule
    & VS-100h & VS-1000h & VS-34kh \\
    \midrule
    SoTA \cite{li2021better} & & & 4.8 \\
    \midrule
    Our Results & 6.9 & 5.0 & \textbf{4.1} \\
    \bottomrule
  \end{tabular}}
\end{table}

\subsection{SpeechStew}

\begin{table*}[t]
\caption{WERs (\%) across multiple tasks for multiple settings compared against pre-existing baselines. We present the performance of the fine-tuned ConformerXXL-P model as well as the result obtained by applying an NST loop starting with this model using the YT-T data as the unlabeled dataset. The result from fine-tuning the PS-model trained upstream with pseudo-labeled YT-T data (see Table \ref{t:ytmain}) are presented in the last row. $^\dagger$Evaluated with punctuation removal following \cite{likhomanenko2020rethinking}. $^\ddagger$Evaluated after removing <unk> tokens. <unk> token removal affects TED-LIUM performance only.}
\centering
\resizebox{0.95\textwidth}{!}{%
\begin{tabular}{lccccccccc}
\toprule
\bfseries Task&\multicolumn{2}{c}{AMI} & Common Voice$^\dagger$ & \multicolumn{2}{c}{LibriSpeech} & \multicolumn{2}{c}{Switchboard/Fisher} & TED-LIUM & WSJ \\
\midrule
 & IHM & SDM1 & & clean & other & SWBD & CH & & eval92 \\
\midrule
\bfseries Prior Work \\
\quad SoTA & 9.0 \cite{speechstew} & 21.2 \cite{kanda2021large} & 8.4 \cite{speechstew} & \textbf{1.4} \cite{ssllimit} & \textbf{2.6} \cite{ssllimit} & \textbf{4.3} \cite{tuske2021limit} & \textbf{6.8} \cite{tuske2021limit} & 5.2 \cite{likhomanenko2020rethinking} & \textbf{1.3} \cite{speechstew}\\
\quad ConformerXXL-LibriLight \cite{speechstew} & 9.5 & 22.7 & 8.4 & 1.7 & 3.3 & 4.8 & 10.6 & 5.7 & \textbf{1.3} \\
\midrule
\bfseries Our Work \\
\quad ConformerXXL-RNNT-P & 8.6 & \textbf{17.7} & 7.8 & 1.9 & 3.5 & 4.6 & 10.2 & 5.9 & \textbf{1.3} \\
\quad ~+ Downstream NST & \textbf{7.8} & 18.3 & \textbf{7.7} & 1.9 & 3.7 & 4.5 & 8.2 & 5.2 & 1.6 \\
\quad ~~~ (Non-filtered) & (9.8) & (22.7) & (11.5) & (2.9) & (6.6) & (5.2) & (9.1) & (5.0) & (4.0) \\[2pt]
\quad ConformerXXL-RNNT-PS$^\ddagger$ & 8.3 & 19.5 & 8.8 & 2.1 & 4.1 & 4.8 & 8.4 & \textbf{5.0} & 1.6 \\
\bottomrule
\end{tabular}}
\label{t:speechstew}
\end{table*}

\textbf{Data:}
The SpeechStew \cite{speechstew} dataset is assembled by putting together seven public speech corpora---AMI \cite{ami}, Common Voice \cite{cv}, English Broadcast News%
\footnote{Linguistic data consortium (LDC) datasets LDC97S44, LDC97T22, LDC98S71 and LDC98T28.},
LibriSpeech \cite{librispeech}, Switchboard/Fisher%
\footnote{LDC datasets LDC2004T19, LDC2005T19, LDC2004S13, LDC2005S13 and LDC97S62.},
TED-LIUM v3 \cite{rousseau2012ted,hernandez2018ted} and Wall Street Journal\footnote{LDC datasets LDC93S6B and LDC94S13B.}.
All utterances from these datasets are collected and mixed randomly and batched for training---no additional steps are taken regarding balancing and mixing data from disparate datasets. The training, dev and evaluation sets for these datasets are process as in \cite{speechstew}, where the transcripts have been prepared via Kaldi \cite{kaldi}. The inference results on the test sets are scored via corresponding Kaldi scripts, while for the Common Voice set, we take the extra step of dropping punctuation before evaluating the word error rates. We use a 1k-WPM constructed based on the LibriSpeech test set for training P-models, while the PS-models use the upstream 4k-WPM.

\textbf{Overview:}
We have compiled key experimental results for SpeechStew in Table \ref{t:speechstew}. As a baseline, we have listed SoTA word error rates for each task inside SpeechStew, and recorded the performance of the ConformerXXL RNN-T network pre-trained with Libri-Light data trained on SpeechStew \cite{speechstew}.

\textbf{Libri-Light vs. YT-U: } We report the performance of the ConformerXXL-RNNT-P model trained on SpeechStew in the third row of Table \ref{t:speechstew}. As in \cite{speechstew}, the model is trained on the mixed SpeechStew data, without any data balancing or batch-wise mixing. By comparing the second and third rows, we see that pre-training on YT-U is able to benefit a wider variety of domains compared to LibriSpeech. This result is not surprising, since YT-U is a much bigger (1 million vs. 60k hours), diverse (see Section \ref{ss:ytdata}) dataset.

\textbf{Downstream Noisy Student Training: } We experiment with downstream noisy student training, where we apply one NST loop with the ConformerXXL-RNNT-P model trained on SpeechStew. To do so, we pseudo-label the YT-T dataset with the SpeechStew-trained Conformer model, and filter 50\% of the data using confidence-per-word of the generated transcripts. We then mix SpeechStew data with the pseudo-labeled data without any balancing. The result is recorded in the fourth row of Table \ref{t:speechstew}, where we see significant improvement in the AMI-IHM, Callhome and TED-LIUM test sets, with small performance degradation on other test sets. We find filtering to be a crucial part of NST. Comparing the performance of models trained on pseudo-labeled datasets constructed with and without filtering (fourth vs. fifth row of Table \ref{t:speechstew}), we find that without filtering, training with the pseudo-labeled data lead to severe degradation of performance across the board, save for a surprising performance improvement on TED-LIUM.

\textbf{PS-models: } We have also trained the ConformerXXL-RNNT-PS on the SpeechStew dataset. Deviating from the general trend observed in the rest of the paper, the PS-model does worse than the P-model across the board as can be observed by comparing the third and last rows of Table \ref{t:speechstew}. We hypothesize the performance lag of the PS-model comes in part from differences in the text normalization of upstream and downstream labeled tasks. In contrast to PS-models, P-models can use WPMs native to the downstream task, and learn native text normalization conventions from the beginning. A manifestation of this disadvantage we have observed is that the fine-tuned PS-model has a tendency to produce <unk> tokens during pauses, resulting in showing 8.7\% WER on the TED-LIUM test set---simply getting rid of these tokens led to a 3.7\% absolute improvement in WER. This issue did not affect any of the other test set performances. 

\textbf{LibriSpeech and Switchboard/Fisher: } The performance of YT-U pre-trained models lag behind SoTA performance on the LibriSpeech and Switchboard/Fisher tasks. There are two main factors that contribute to this discrepancy. For both LibriSpeech \cite{ssllimit} and Switchboard \cite{tuske2021limit}, language models trained on text corpora coordinated for each task has been used to improve the ASR performance, while no language models have been employed for our results. For LibriSpeech, the pre-training dataset Libri-Light \cite{librilight}, which specifically matches the domain of the LibriSpeech corpus, has been utilized for pre-training and self-training for achieving SoTA in \cite{ssllimit}.

\subsection{CHiME-6}

\textbf{Data:}
CHiME-6 \cite{chime6} contains 40 hours of distant microphone conversational speech recognition in everyday home environments. We use the official front-end enhancement recipe \cite{chime6} to enhance the dataset---BeamformIt was used to create an augmented training set, while guided source separation \cite{Boeddecker2018} with 12 channels was used to enhance the dev/evaluation sets.

\textbf{Overview:}
We have presented results from training P- and PS-models on CHiME-6 in Table \ref{t:chime6}. As baselines, we have listed the performances of previous SOTA models and Libri-Light pre-trained models reported in the literature.

\begin{table}[h]
\caption{WERs (\%) on CHiME-6. We show the dataset used for pre-training and upstream labeled-training before fine-tuning the model on the CHiME-6 dataset. Recall that PS-models are trained upstream on the pseudo-labeled YT-T dataset.}
\centering
\resizebox{0.95\columnwidth}{!}{%
\begin{tabular}{lccrr}
\toprule
Model & Pre-training & Upstream & Dev & Eval \\
\midrule
\textbf{Prior Work} \\
\quad HMM Baseline \cite{chime6} & - & - & 51.8 & 51.3 \\
\quad HMM (SOTA) \cite{medennikov2020stc} & - & - & 36.9 & 38.6 \\
\quad ConformerXXL \cite{speechstew} & Libri-Light & - & - & - \\
\quad ConformerXXL \cite{speechstew} & Libri-Light & SpeechStew & 31.9 & 38.9 \\
\midrule
\textbf{Our Work} \\
\quad ConformerXXL-RNNT-P & YT-U & - & 35.1 & 39.5 \\
\quad ConformerXXL-RNNT-P & YT-U & SpeechStew & \textbf{26.2} & 34.4 \\
\quad ConformerXXL-RNNT-PS & YT-U & YT-T & \textbf{26.2} & \textbf{31.0} \\
\bottomrule
\end{tabular}}
\label{t:chime6}
\end{table}

\textbf{Results: } We have recorded the performance of the ConformerXXL-RNNT-P model directly trained on CHiME-6 in the third row of Table \ref{t:chime6}, while that of the model first trained on SpeechStew and fine-tuned on CHiME-6 is recorded in the fourth row. Quite surprisingly, the P-model directly trained on CHiME-6 shows strong performance, in contrast to the Conformer XXL pre-trained with Libri-Light, which fails to directly train on CHiME-6. Upon training on the upstream task of SpeechStew and fine-tuning on CHiME-6, we are able to exceed state-of-the-art performance with 11\% relative WER improvement on the CHiME-6 test set. The strongest CHiME-6 performance is achieved by fine-tuning the PS-model, resulting in a 20\% relative WER improvement on the test set.

\subsection{Telephony}

\textbf{Data:}
We aim to improve an English (GB) telephony task with a training set obtained by mixing a labeled telephony dataset with 320 hours of audio and a video-based dataset with 30 hours of audio. We use two test sets to evaluate the model. {\it Test-short} consist of 9 hours of telephony audio, while {\it Test-long} consist of 82 hours of long video-based audio. The two test sets are chosen because we wish to construct a model that performs well on telephony audio, but at the same time be able to show good performance on long utterances.

\textbf{Overview:}
Our objective will be two-fold. The first goal will be to train a large ASR model that does well on the Telephony task, while the next will be to distill its performance to a streaming model. As the base streaming model, we use the RNN-T model of \cite{he2019streaming} with an 8-layer uni-directional LSTM decoder with cell size 2048, and a 2-layer LSTM decoder with the same cell size trained with the large multi-domain dataset presented in \cite{multidomain}. As a baseline, we fine-tune this model with various mixtures of the telephony dataset and the video dataset. Our results are summarized in Table \ref{t:telephony}.

\begin{table}[h]
\caption{Telephony test WERs (\%). The performance from fine-tuning the P- and PS-models are presented. Note that the PS-model, even before seeing the Telephony data, performs reasonably on the task. The fine-tuned PS-model is used to generate NST data for training the student streaming model.}
\vskip 0.1in
\centering
\resizebox{0.98\columnwidth}{!}{%
\begin{tabular}{lcccrr}
\toprule
Model & \multicolumn{3}{c}{Fine-tuning Mixture} & Test-short & Test-long \\
\cmidrule{2-4}
& Telephony & Video & NST & \\
\midrule
\textbf{Streaming Model} \\
\quad Baseline & N/A & N/A & N/A & 33.11 & 15.53 \\
\quad Fine-tuned & 1.0 & - & N/A & 22.45 & 21.41 \\
& 0.8 & 0.2 & N/A & 22.64 & 19.99 \\
\midrule
\textbf{ConformerXL-RNNT-P} & 0.8 & 0.2 & N/A & 22.24 & 14.55 \\
\midrule
\textbf{ConformerXL-RNNT-PS} \\
\quad Baseline & N/A & N/A & N/A & 27.20 & 10.97 \\
\quad Fine-tuned & 0.8 & 0.2 & N/A & \textbf{21.24} & \textbf{10.72} \\
\midrule
\textbf{Student Streaming Model} & 0.8 & - & 0.2 & 22.97 & 16.75 \\
\bottomrule
\end{tabular}}
\label{t:telephony}
\end{table}

\textbf{Non-streaming Models: }
We are able to obtain models that perform better than the fine-tuned streaming models on both tasks by training the ConformerXL-RNNT-P and PS models. The PS-model exhibits improved performance on Test-short, while it is able to lower the Test-long WER by a significant amount compared to any of the streaming baselines.

\textbf{Distilling to Streaming Models: }
We attempt to distill the performance of the fine-tuned PS-model to the streaming model by taking a random 20\% subset of YT-U and pseudo-labeling it with the model. We apply 50\% filtering according to confidence-per-word to generate the NST dataset that is in turn used to train the streaming model. The result of mixing this data with the labeled data to fine-tune the streaming model is given in the last row of Table \ref{t:telephony}. We are able to improve the Test-long performance of the streaming model by a relative 16\% while suffering a minuscule performance loss on Test-short.

\section{Experiments with Voice Search}
\label{s:VS}

We now move on to conduct a series of experiments to explore the effect of pre-training, upstream and downstream self-training and model size scaling for downstream tasks of different scales. To conduct a systematic study, especially with respect to the effect of the scale of the downstream task, we choose to study the Voice Search task \cite{li2021scaling}, which has a large amount of labeled audio, and produce tasks of varying scale by sub-sampling. By doing so we find that the combination of increasing the model size and utilizing a large unlabeled dataset vastly improves labeled-data efficiency.

\subsection{Data}
\label{ss:searchdata}

\textbf{English (US):}
Our principal dataset is the English (US) Voice Search dataset, containing 34k hours of labeled voice search audio  \cite{li2021scaling}. As noted before, we sample random 1000 and 100 hour subsets, VS-1000h and VS-100h. The transcripts for the audio are tokenized either using graphemes or a 4k-token WPM. The Conformer G models are trained using WPM tokenization while grapheme tokenization is used for experiments with the XL/XXL model unless indicated otherwise. A 128M-parameter Conformer LM trained on a large corpus of in-domain text for improving the performance of the ASR models further.

\textbf{Non-English:}
To explore cross-lingual benefits of pre-training, we examine three Voice Search tasks in non-English languages---Hungarian (HU), Chinese (TW) and Hindi (IN) \cite{li2021scaling}. For each language we prepare an unlabeled YouTube dataset segmented using voice activation detection (VAD \cite{zazo2016feature}), the labeled Voice Search dataset and its 100h and 1000h subsets. The amount of unlabeled YouTube data and labeled Voice Search data are tabulated in Table \ref{t:nonenglishdata}.

\begin{table}[h]
\caption{YouTube and Voice Search datasets.}
\centering
\resizebox{0.8\columnwidth}{!}{%
\begin{tabular}{ccc}
\toprule
Language & YouTube (hrs) & Voice Search (hrs) \\
\midrule
Hungarian (HU) & 400k & 9k \\
Chinese (TW) & 900k & 20k \\
Hindi (IN) & 800k & 27k \\
\bottomrule
\end{tabular}}
\label{t:nonenglishdata}
\end{table}

\subsection{Pre-training}

We compare the results from training the ConformerXL-RNNT from scratch on the VS-100h, 1000h and 34kh sets against training the pre-trained model ConformerXL-RNNT-P on these tasks. The results, plotted in the first panel of Figure \ref{f:returns} in yellow and green, illustrate the benefits of pre-training. We find that while the gains from utilizing additional unlabeled and labeled data exists even at very large downstream dataset size, the relative improvement in performance decreases.

\subsection{Scaling-up Model Size}

We train RNNT-P models of three different sizes, XL, XXL and G on the VS-100h, 1000h and 34kh datasets. The 8B parameter G models are trained using WPM tokenization. The results are plotted in the second panel of Figure \ref{f:returns}.

\begin{figure}[h!]
\centering
\includegraphics[width=0.78\columnwidth]{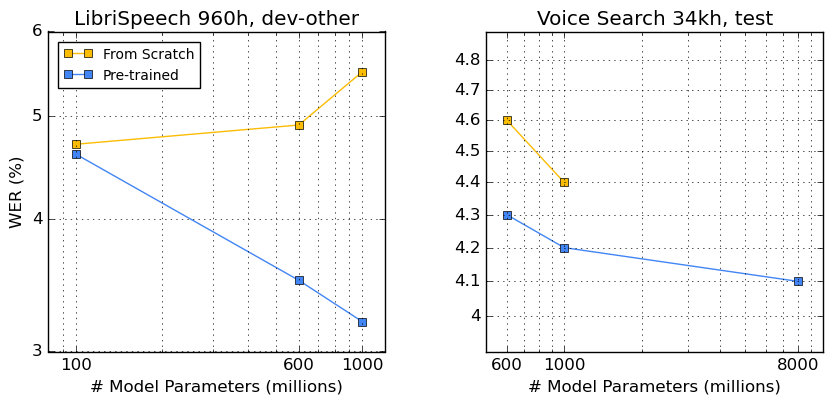}
\caption{\textbf{(Left)} The LibriSpeech dev-other performance of Conformer models of varying size and pre-training conditions when trained on LibriSpeech 960h reported in \cite{ssllimit}. \textbf{(Right)} Voice Search 34kh test performance of Conformer models.}
\label{f:largept}
\end{figure}

Meanwhile, we address the phenomenon observed in \cite{ssllimit}, where it was shown that for LibriSpeech, larger Conformer models performed worse unless they are pre-trained. Their results are plotted in the first panel of Figure \ref{f:largept}.

To see if this pattern still holds for very large labeled datasets, we train our Conformers from scratch on the entirety of Voice Search and compare their performances with those of their pre-trained counterparts. We find that in our case, the trend of pre-training being necessary for benefiting from model size does not hold anymore in the 600M to 1B parameter range as is shown in the second panel of Figure \ref{f:largept}. The 8B parameter model training fails to converge for this task.

\subsection{Cross-lingual Benefits}

We explore cross-lingual benefits of pre-training by examining Voice Search tasks in Hungarian (HU), Chinese (TW) and Hindi (IN). For each language, we prepare three Conformer XL RNN-T models: a baseline model with no pre-training, a model pre-trained with English YouTube data and a model pre-trained with YouTube data in the native language. We train each model on the entire Voice Search set and its 100h and 1000h subsets. To make a fair comparison between cross-lingual pre-training and native pre-training, rather than using P-models trained on YT-U, we prepare an unlabeled English YouTube dataset also segmented by a VAD \cite{zazo2016feature} with 926k hours of audio, and pre-train our models on this dataset.

\begin{figure}[h!]
\centering
\includegraphics[width=0.98\columnwidth]{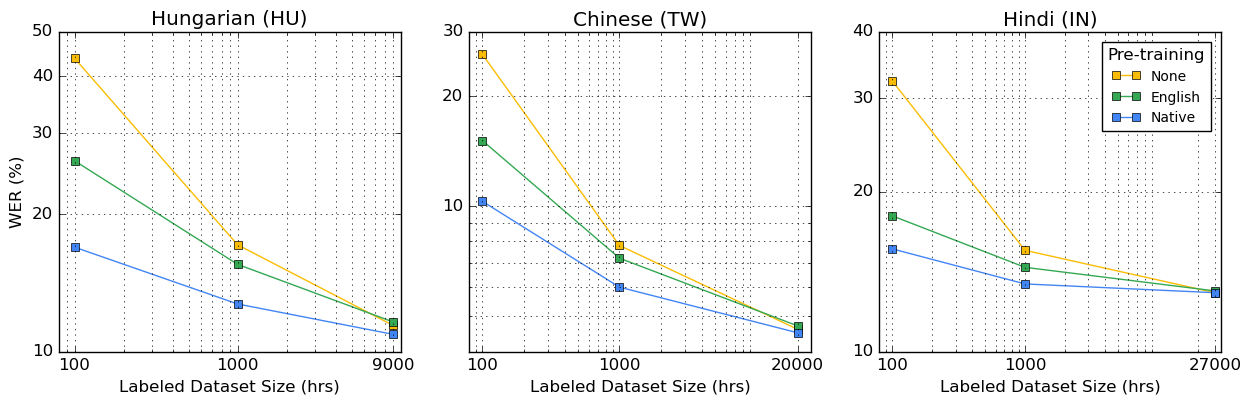}
\caption{Test WERs (\%) from training pre-trained Conformer XL RNN-T networks on non-English Search datasets and subsets thereof. Both axes are plotted in log-scale.}
\label{f:nonenglish}
\end{figure}

The results of the experiments are plotted in Figure \ref{f:nonenglish}. Consistent with the overall theme of this section, both English and native pre-training are more effective with smaller labeled dataset size. While we find cross-lingual benefits of pre-training at 100h and 1000h labeled datasets, we can see English pre-training hurting the performance when training on the full dataset. We can observe, however, that the benefit from native pre-training persisting up to full dataset size for Hungarian (HU) and Chinese (TW).

\subsection{Upstream Self-training}

We fine-tune the ConformerXL-RNNT-PS model, pre-trained with YT-U and self-trained upstream on the YT-T dataset, on the VS-100h, 1000h and 34kh sets. These results, plotted in blue in the first panel of Figure \ref{f:returns}, show further gains beyond the pre-trained model for the 100h and 1000h training sets compared to the baselines trained from scratch. The fine-tuned PS-model is not able to achieve better performance than the fine-tuned P-model when trained on the full VS dataset. The effect of upstream self-training on the XXL and G models remains to be investigated.

\subsection{Downstream Self-training and LM Fusion}
\label{ss:searchnst}

We apply downstream noisy student training to VS-100h and 1000h with RNNT-P models. To do so, a teacher model is trained on the same labeled dataset and generates pseudo-labels for a random 20\% subset of YT-U to train a student model with. 50\% of the teacher-generated transcripts are filtered out based on confidence-per-word and mixed batch-wise with the labeled data. To further experiment with language model fusion, we use WPM tokenization for training the student models.

\begin{table}[h]
\caption{Settings for downstream NST experiments.}
\centering
\resizebox{0.98\columnwidth}{!}{%
\begin{tabular}{cccccc}
\toprule
Experiment & Teacher Model & Teacher Tokens & Teacher WER & NST Ratio\\
\midrule
XL on 1000h & XL-CTC-P & WPM & 8.0\% & 0.6\\
G on 100h & XXL-RNNT-P & Grapheme & 9.4\% & 0.4 \\
G on 1000h & XXL-RNNT-P & Grapheme & 6.2\% & 0.6 \\
\bottomrule
\end{tabular}}
\label{t:nstsetting}
\end{table}

We have conducted three experiments, two where the ConformerG-RNNT-P is used as a student for the 100h/1000h tasks, and one where the ConformerXL is used as a student for the 1000h task. Some relevant information for these models is listed in Table \ref{t:nstsetting}. Note that we have used a CTC instead of an RNN-T network as a teacher for the XL model so that its architecture does not exactly coincide with that of the student. The NST ratio indicates the ratio of teacher-generated transcripts within the student training batch.

\begin{table}[h]
\caption{Voice Search test WERs (\%) for Conformer RNNT-P models with and without NST and LM fusion.}
\centering
\resizebox{0.98\columnwidth}{!}{%
\begin{tabular}{lrrcrrcrr}
\toprule
& \multicolumn{2}{c}{G on 100h} &&
\multicolumn{2}{c}{XL on 1000h} &&
\multicolumn{2}{c}{G on 1000h}   \\
\midrule
& No LM & LM && No LM & LM && No LM & LM\\
\midrule
Baseline & 8.8 & 7.7 && 6.5 & 6.1 && 5.5 & \textbf{5.0}\\
+ NST & 7.8 & \textbf{6.9} && 6.0 & \textbf{5.5} && 5.3 & \textbf{5.0}\\
\bottomrule
\end{tabular}}
\label{t:searchnst}
\end{table}

As noted in Section \ref{ss:searchdata}, we utilize a Conformer language model for shallow fusion \cite{shallowfusion} with the trained ASR models. The fusion weight \cite{shallowfusion} and the non-blank reward \cite{sak2015fast,zhang2020transformer} are selected by a small random exploration. The result of applying NST and LM fusion is given in Table \ref{t:searchnst}.

We have not been able to achieve additional gains on the full Voice Search task using downstream NST. Some discussion on this matter is given in the final section.

\section{Non-ASR Tasks}
\label{s:downstreamnonasr}

We now explore the utility of the representations of pre-trained Conformers for audio classification tasks. In this section, we consider the Conformer XL Non-RA, a Conformer XL model that does not use relative attention \cite{dai2019transformer}, which turns out to outperform its relative attention counterpart on these tasks.

\subsection{Non-Semantic Speech (NOSS) Benchmark}
\label{ss:noss}

\textbf{Tasks and Datasets:}
The Non-Semantic Speech Benchmark (NOSS) \cite{trill} is a benchmark of speech classification tasks that is used to compare the usefulness of speech representations. The benchmark includes a variety of tasks such as speech emotion recognition \cite{cremad, savee}, speaker identification \cite{voxceleb}, and language identification \cite{voxforge}, but specifically excludes tasks that are focused on the meaning of words. We follow \cite{frill} and include three health-speech tasks in our representation evaluation: mask detection during speech \cite{compare2020}, environmental human sound detection such as coughing and sneezing \cite{esc50} and dementia detection \cite{boller2005dementiabank}. Table \ref{tab:noss} describes the datasets.

\begin{table}[h!]
\centering
\caption{Non-Semantic Speech Benchmark (NOSS) datasets. Average duration is in seconds. $^\dagger$This is a subset of VoxCeleb filtered according to YouTube's privacy guidelines. $^\ddagger$This is a subset of ESC-50 with human sound labels.}
\resizebox{0.98\columnwidth}{!}{
\begin{tabular}{ c c c c c} \toprule
  Dataset & Target & Classes & Samples &  Avg duration (s) \\
  \midrule
 VoxCeleb$^\dagger$~\cite{voxceleb} & Speaker id & 1,251 & 12,052 & 8.4 \\
 VoxForge~\cite{voxforge} & Language id & 6 & 176,438 & 5.8 \\
 Speech Commands~\cite{speechcommands} & Command & 12 & 100,503 & 1.0 \\
 CREMA-D~\cite{cremad} & Emotion & 6 & 7,438 & 2.5 \\
 SAVEE~\cite{savee} & Emotion & 7 & 480 & 3.8 \\
 Masked Speech~\cite{compare2020} & Mask wearing & 2 & 36,554 & 1.0   \\
 ESC-50 human$^\ddagger$~\cite{esc50} & Non-speech sounds & 10 & 386 & 13.8 \\
 DementiaBank~\cite{boller2005dementiabank} & Dementia/healthy & 2 & 210 & 70.0 \\
 \bottomrule
\end{tabular}}
\label{tab:noss}
\end{table}

\begin{table*}[t!]
\centering
\caption{Accuracies (\%) for NOSS tasks. $^\dagger$A filtered subset of VoxCeleb1 according to YouTube’s privacy guidelines has been used. $^\ddagger$The Masked Speech task performance is reported using unweighted average recall~\cite{compare2020} instead of accuracy. $^*$Audio and visual features used. $^{**}$Acoustic and textual features used. $^\mathsection$Layer 10 is used as in \cite{trill}.}
\vskip 0.1in
\label{tab:best_models}
\resizebox{0.95\textwidth}{!}{
\begin{tabular}{lcccccccc}
\toprule
Model & VoxCeleb1$^\dagger$ & Voxforge & \begin{tabular}{@{}c@{}}Speech \\ Commands\end{tabular}   &  CREMA-D & SAVEE & \begin{tabular}{@{}c@{}}Masked$^\ddagger$ \\ Speech\end{tabular}  & \begin{tabular}{@{}c@{}}ESC-50 \\ HS\end{tabular} & DementiaBank\\
\midrule
\textbf{Previous SoTA}
& - & 95.4 \cite{sarthak2019spoken} & \textbf{97.9}~\cite{speech_commandssota} & 74.0$^*$ \cite{ghaleb2019} & 70.0 \cite{frill} & 73.0~\cite{Szep2020} & \textbf{93.9} \cite{trill} & \textbf{80.6}$^{**}$ \cite{db2017}  \\
\midrule
\textbf{Baselines} \\
\quad TRILL \cite{trill}
& 13.1 & 84.5 & 77.6 & 65.8 & 65.0 & 65.3 & 86.4 & 63.7 \\
\quad FRILL \cite{frill}
& 13.8 & 78.8 & 74.4 & 71.3 & 63.3 & 67.2 & 87.9 & 68.7 \\
\quad YAMNet$^\mathsection$ \cite{yamnet}
& 9.6 & 79.8 & 78.5 & 66.4 & 69.2 & 59.6 & \textbf{93.9} & 62.7 \\
\quad ASR Encoder \cite{asremb}
& 5.2 & 98.9 & 96.1 & 71.8 & 85.0 & 54.4 & 75.8 & 63.7 \\
\midrule
\textbf{Our Results} \\
\quad ConformerXL-P
& 49.4 & 99.7 & 95.2 & 86.8 & \textbf{92.5} & 68.0 & 89.4 & 60.8 \\
\quad ConformerXL-P Non-RA
& 50.3 & 99.7 & 97.5 & \textbf{88.2} & \textbf{92.5} & \textbf{73.4} & 89.4 & 72.5 \\
\quad ConformerXXL-P
& \textbf{53.3} & 99.6 & 96.3 & 85.5 & 87.5 & 68.6 & 90.9 & 65.7 \\
\quad ConformerG-P
& 48.9 & \textbf{99.8} & 90.1 & 87.1 & 90.0 & 61.3 & 70.3 & 54.9 \\
\bottomrule
\end{tabular}}
\label{table:noss}
\end{table*}

\textbf{Evaluation:}
We modify the evaluation method described in \cite{trill}. For every (model, layer, task) triplet, we train three types linear models using the Scikit-Learn library \cite{sklearn} (logistic regression, balanced logistic regression, linear discriminant analysis). Given the model and task, we select the layer and regression method that yields the maximum {dev-set} performance, and report the {test-set} performance of that configuration in table~\ref{table:noss} against previously reported state-of-the-art results. In the second block of rows in the table, we present the results of baseline NOSS task evaluations we have conducted using audio representations obtained by methods previously studied in the literature \cite{trill, frill, asremb, yamnet}.

\begin{itemize}
\item We achieve new SoTA on 4/7 public tasks%
\footnote{We exclude our VoxCeleb results from this group, since we use a filtered subset that does not have associated public benchmarks.}
using only a task-specific linear layer (see table~\ref{tab:noss}). Linear models trained on the Conformer embeddings outperform previously reported results on Voxforge, CREMA-D, SAVEE and Masked Speech that were achieved by using complex, task-specific architecture and training. 

\item Conformers on acoustic features alone are competitive with other models that use multimodal data. In particular, we achieve 20\% relative improvement on the previous SoTA CREMA-D result obtained using visual and acoustic features. Our results are worse than SoTA on DementiaBank, which has been achieved by using additional textual information along with the acoustic features \cite{db2017}.

\item The Conformer XL without relative attention produces the best-performing features (see Table \ref{tab:noss}) consistently. This is also confirmed by the analysis using the average accuracy measure presented below.

\item On most tasks, all four Conformer models outperform all previous SoTA numbers from the other non-semantic speech representations ("Baseline" rows in Table \ref{table:noss}), with the exception of the ESC-50 performance of YAMNet, whose labeled classes of the supervised training set is a super-set of that of ESC-50.
\end{itemize}

\begin{figure}[h!]
  \centering
  \includegraphics[width=0.90\columnwidth]{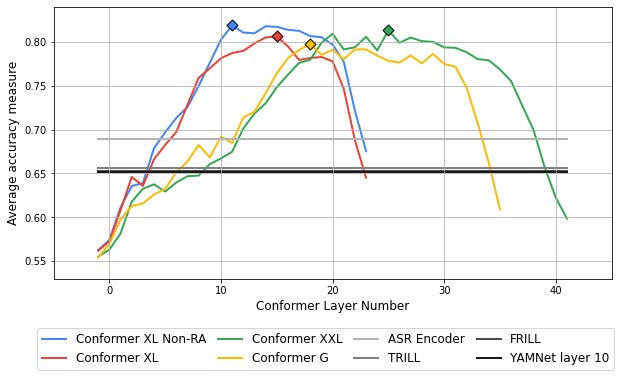}
  \caption{Average accuracy measure for each Conformer model and layer. The model's best layer is marked. The average accuracy measure for ASR Encoder~\cite{asremb}, TRILL~\cite{trill}, FRILL~\cite{frill}, and YAMNet layer 10~\cite{yamnet} are marked as horizontal lines for reference.}
  \label{fig:noss_context}
\end{figure}

\textbf{Average Accuracy Measure:}
As in \cite{frill}, we use the average accuracy measure over all NOSS tasks as metric for quantifying the general quality of an embedding. In figure~\ref{fig:noss_context}, we plot the average accuracy measure as a function of the model layer, starting from the positional embedding layer indexed as layer -1. The curves for the Conformer models truncate at the penultimate layer of the model.

\begin{itemize}
\item We find that embeddings from the majority of Conformer layers outperform previous results. This suggests the general usefulness of the representations within the pre-trained Conformer models for non-semantic speech tasks, not just tasks related to language.
\item The best Conformer layer is in the middle of the network, not at the penultimate position. This suggests that even though the Conformer learned a new SoTA non-semantic speech representation, the training procedure could be modified for further improvement.
\end{itemize}

\subsection{AudioSet}
\label{ss:audioset}

\textbf{Data and Overview:}
In the preceding sections, we have demonstrated the capability of our Conformer models in a wide variety of speech processing tasks.  We now evaluate the utility of our unsupervised representation for the \emph{non-speech} task of audio event classification.  For this, we consider the commonly used AudioSet benchmark~\cite{gemmeke2017audio}, which includes nearly 2 million audio clips, each with one or more labels drawn from an ontology of 527 classes.  More specifically, we use the unsupervised representation evaluation protocol used by several recent papers.  This involves 1) using our pre-trained Conformer models to extract a fixed-dimensional frame-level representation for each clip; 2) using AudioSet to train a simple frame-level classifier on top consisting of a single hidden layer with 512 ReLU units followed by a linear classification layer with 527 independent sigmoid outputs; 3) applying this to each evaluation clip and mean pooling the frame-level scores to report an overall mean average precision value.  Past approaches that have used this evaluation paradigm have used the same AudioSet training data for unsupervised pre-training.  Thus, in addition to reporting performance using our models pre-trained with YT-U, we also pre-trained an additional ConformerXL with AudioSet to limit domain mismatch and make results more comparable to past work. Table~\ref{t:audioset} compares the AudioSet classification performance for several past approaches and various representations derived from the pre-trained ConformerXL (both with and without relative attention).  For each Conformer model, we report both the performance at the output (layer 24) and at the best internal Conformer layer as determined using a small portion of AudioSet train held out as a development set.  

\begin{table}[h]
\caption{AudioSet shallow classifier benchmark results in mean average precision (mAP). Following standard practice for evaluating unsupervised representations~\cite{jansen2018unsupervised}, the representation is held fixed and the entire AudioSet train set is used to train an MLP classifier with a single 512-unit hidden layer. Some prior work uses both audio spectrograms (A) and video (V) as inputs during training where indicated. All results shown only use spectrograms as input for evaluation.}
\vskip 0.1in
\centering
\resizebox{0.95\columnwidth}{!}{
\begin{tabular}{lccc}
\toprule
Representation & Pre-training & Modalities & mAP \\
\midrule
\textbf{Previous SoTA} \\
\quad Multi-format~\cite{wang2021multi} & AudioSet & A & 0.329 \\
\quad BRaVE~\cite{recasens2021broaden} & AudioSet & A,V & 0.347 \\
\midrule
\textbf{ConformerXL} \\
\quad Output Layer & YT-U & A & 0.192 \\
\quad Best Layer (10) & YT-U & A & 0.304 \\
\textbf{ConformerXL Non-RA} \\
\quad {Output Layer} & YT-U & A & 0.234 \\
                                    & AudioSet & A & 0.222 \\
\quad {Best-Dev Layer (10)} & YT-U & A & 0.308 \\
                                       & AudioSet & A & 0.340 \\
\bottomrule
\end{tabular}}
\label{t:audioset}
\end{table}

\textbf{Results:}
We can make several conclusions from these results.  First, we find that while the representation defined by the final Conformer output layer lags all past approaches, selecting internal layers performs significantly better, as was the case for the NOSS experiments above.  This indicates the wav2vec 2.0 pretraining objective is not well matched to the audio event classification task, but that it still induces useful internal representations.  Second, relative attention results in a small degradation of performance for this task, which is in line with NOSS and counter to the ASR results.  Like the NOSS tasks, the AudioSet benchmark involves whole-clip level predictions, which we hypothesize likely reduces the value of relative positional information.   Third, we find that even when pretraining on speech alone (YT-U), we still learn a representation at layer 10 that performs respectably on the AudioSet benchmark (0.308 mAP) compared with past approaches, even though this model has not been exposed to the diversity of the AudioSet ontology.  However, when we retrain ConformerXL using in-domain AudioSet data, we achieve additional improvement.  Our AudioSet-pre-trained ConformerXL performance of 0.340 mAP outpaces all past work that used spectrograms alone for training and evaluation.  Most past unimodal approaches rely heavily on augmentation in the learning procedures, the types of which need to be carefully chosen to optimize for this task.  However, the Conformer model obtains strong performance without using augmentation of any kind, instead driven solely by the architectural design and pre-training objective.

\section{Discussion and Future Directions}

\subsection{ASR Model and Data Efficiency}

\textbf{Data Efficiency:}
We find that increasing model size, pre-training, upstream and downstream self training positively affects the training performance, but mainly in labeled data efficiency. When the labeled dataset size grows very large, the effect of many of these methods become smaller.

\textbf{Pre-training and Training Stability for Large Models:}
We find that training on labeled datasets becomes harder and more brittle as the model gets very large. The problem is exacerbated by the training cost of the larger models, making hyperparameter tuning prohibitively expensive. As a result, our 8B model becomes practically un-trainable on the full Voice Search dataset. Large model training becomes stable after they are pre-trained.

\textbf{Model Compression:}
A natural direction of research that emerges from this work is to find ways to practically benefit from the performance gains achieved by giant models. Research on how to compress giant models with minimal performance loss will be a crucial element for such endeavors.

\subsection{Downstream ASR Tasks}

\textbf{P- or PS-models?:}
Since PS-models utilize additional upstream labeled data, it should be expected to perform better than P-models on small downstream tasks in general, which is what we find. Meanwhile, there are factors that make the PS-models behave worse in the approach taken in this work, where we aim to train a universal upstream model PS that is then fine-tuned on many downstream tasks. In this work, the PS-models are trained in a text environment (e.g., text normalization, tokenization) optimized for the upstream task, while the P-models can be trained on an environment native to the downstream task from the beginning, which might differ significantly from the upstream task. Furthermore, difference in these factors (e.g., tokenization) can make it unwieldy to use available downstream resources (e.g., LM fusion).

\textbf{Downstream NST training for full Voice Search sets:}
Application of the standard downstream NST recipe used in this paper of pseudo-labeling either YT-U or YT-T, filtering 50\% using confidence-per-word and mixing the pseudo-labeled data with labeled data, has not been able to improve full Voice Search set performance. Our experiments, which have been conducted with the English (US) and Hungarian (HU) full Voice Search set, have in fact lead to slight degradation of performance. A tentative conclusion can be that pseudo-labeling becomes less effective when the labeled dataset set size becomes very large, although a more through investigation would be needed to sharpen this assertion.

\subsection{Downstream Non-ASR Tasks}

\textbf{Beyond Encoded Features:}
We have restricted the use of large pre-trained models as feature encoders for non-ASR tasks in this work. It would be interesting to go beyond this use to further improve audio classification tasks in general.

\small

\appendices
\section{Experiment Details} \label{ap:details}

\subsection{Pre-training}
Some pre-training parameters are summarized in Table \ref{t:pparams}. Pre-training has been carried out using Google Cloud TPU V3 chips.

\begin{table}[h!]
  \caption{Pre-training parameters.}
  \label{t:pparams}
  \centering
  \resizebox{0.95\columnwidth}{!}{%
  \begin{tabular}{ccccccc}
    \toprule
    Model & Batch Size & \# TPU Cores & Days & Epochs & Warm-up Steps & Peak LR\\
    \midrule
    XL & 4096 & 512 & 5 & 10 & 25k & 1e-3 \\
    XXL & 4096 & 1024 & 8 & 10 & 25k & 1e-3 \\
    G & 1280 & 1024 & 18 & 4 & 50k & 1e-4 \\
    \bottomrule
  \end{tabular}}
\end{table}

\subsection{Voice Search}

\textbf{P-Models:}
When training XL/XXL P-models on VS-100h, 1000h and 34kh, a fixed transformer learning rate schedule is used for all three datasets, while the batch size is scaled up by a factor of 4 for each bigger task. The decoder learning rate schedule has peak learning rate 1e-3 and warm-up steps 1.5k for both models. Meanwhile, the encoder learning rate schedule has peak learning rate 3e-4/2.4e-4 for the XL/XXL models respectively, with 5k warm-up steps. The batch size for the VS-100h task is set to 128.

\textbf{PS-Models:}
For fine-tuning XL PS-models on VS-100h, 1000h and 34kh, we use a transformer learning rate schedule with fixed warm-up step-size, but adjust the learning rate for the three tasks. The batch size is scaled up by a factor of 4 for each bigger task, while the learning rate is scaled up by a factor of 3 accordingly. For the PS-models, the encoder and decoder learning rate schedule is set to be the same. The VS-100h task learning rate schedule has peak learning rate 3e-5 and 5k warm-up steps. The batch size is set to 128.

\textbf{Training from Scratch:} The Conformer XL and XXL models trained from scratch have been trained with batch size 1024 using a transformer learning rate schedule with peak learning rate 1.8e-3/3.5e-4 respectively and 33k warm-up steps.

\textbf{Noisy Student Training:}
Training parameters when training with the combined supervised and teacher generated data are kept the same as supervised training. We find the model performance to improve over a longer period of time, and require 2x to 4x training time compared to supervised training for convergence.

\subsection{Public Datasets}

\textbf{SpeechStew: } The hyperparameters used for training the 1B-parameter P-model on the SpeechStew data are equivalent to those given in section 3 of \cite{speechstew} for training their 1B-parameter model pre-trained on Libri-Light. In particular, the supervised training is carried out for 100k steps with batch size 2048. For noisy student training, the model is trained for 200k steps with batch size 1024.

\textbf{CHiME-6: } The hyperparameters used for training the 1B-parameter PS-model on the CHiME-6 data are equivalent to those given in section 3.1 of \cite{speechstew} for training their 1B-parameter model pre-trained on Libri-Light and trained upstream on SpeechStew. 

\section*{Acknowledgments}
We would like to thank Daniel Adiwardana, Tony Bruguier, Yuan Cao, Zhehuai Chen, Mike Chrzanowski, Alexis Conneau, Xiangyu Dong, Thibault Doutre, Peter Gavin, Blake Hechtman, Ye Jia, Guangda Lai, Benjamin Lee, Chris Lee, Thang Luong, Andy Ly, Marcello Maggioni, Ananya Misra, Erica Moreira, Mohammad Norouzi, Tayo Oguntebi, Bramandia Ramadhana, Andrew Rosenberg, Ruoxin Sang, Jonathan Shen, Trevor Strohman, Weiran Wang, Haoyu Zhang and Yazhou Zu for useful discussions. We also thank Claire Cui and Johan Schalkwyk for their support of this work.

\ifCLASSOPTIONcaptionsoff
  \newpage
\fi

\bibliographystyle{IEEEtran}
\bibliography{references}

\begin{IEEEbiographynophoto}{}
\end{IEEEbiographynophoto}

\end{document}